\documentclass[a4paper,fleqn]{cas-sc}

\usepackage[numbers,sort&compress]{natbib}
\usepackage{lipsum}
\usepackage{subcaption}
\usepackage{amsmath}
\usepackage{amsfonts}
\usepackage{upgreek}
\usepackage{textcomp}
\usepackage{bm}
\usepackage{float}
\usepackage{lineno}
\begin{document}

\let\WriteBookmarks\relax
\def\floatpagepagefraction{1}
\def\textpagefraction{.001}

\def\journalname{Construction and Building Materials}


\shorttitle{Hierarchical meta-modelling of stone injected with CNT/cement mortar}    

\shortauthors{Rodr\'{i}guez \textit{et al.}} 

\title[mode = title]{Hierarchical meta-modelling for fast prediction of the elastic properties of stone injected with CNT/cement mortar}  

\author[label1]{Rub\'{e}n Rodr\'{i}guez-Romero}[orcid=0000-0002-7219-249X]
\ead{rrodriguezr@us.es}

\author[label2]{V\'{i}ctor Comp\'{a}n}[orcid=0000-0002-6478-1833]
\ead{compan@us.es}

\author[label1]{Andr\'{e}s S\'{a}ez}[orcid=000-0001-5734-6238]
\ead{andres@us.es}

\author[label3]{Enrique Garc\'{i}a-Mac\'{i}as}[orcid=0000-0001-5557-144X]
\ead{enriquegm@ugr.es}
\cormark[1]

\affiliation[label1]{addressline={Department of Continuum Mechanics and Structural Analysis, University of Seville, Camino de los Descubrimientos s/n, 41092 Seville, Spain.}}

\affiliation[label2]{addressline={Department of Continuum Mechanics, University of Seville, Avenida Reina Mercedes, 41012 Seville, Spain.}}

\affiliation[label3]{addressline={Department of Structural Mechanics and Hydraulic Engineering, University of Granada,  C/ Dr. Severo Ochoa s/n, 18071 Granada, Spain.}}

\cortext[cor1]{Corresponding author - Department of Structural Mechanics and Hydraulic Engineering, University of Granada, Av. Fuentenueva sn, 18002 Granada, Spain. Phone: +34 958241000, Ext. 20668.}


\begin{abstract}
This paper presents an innovative hierarchical meta-modelling approach for predicting the effective elastic properties of stone injected with composite cement grouts containing Carbon NanoTubes (CNTs) for structural rehabilitation purposes. The study first addresses the homogenization of the nano- and micro-scales through numerical representative volume elements. The first numerical model focuses on the cement matrix doped with randomly oriented CNTs, while the second model represents the porous stone with CNT/cement grout filling the pores. However, the computational burden associated with these models poses a significant limitation when analysing large-scale macrostructural elements. To overcome this challenge, a hierarchical meta-modelling approach based on two nested Kriging surrogate models is proposed. This approach offers a fast and accurate alternative to bypass the time-consuming numerical homogenization process. The first surrogate model establishes a connection between the microstructural characteristics of the CNTs and cement and the effective elastic properties of the composite grout. Subsequently, a second meta-model expands over the first one by introducing two additional variables: the elastic modulus of the stone and the porosity, allowing for estimation of the overall elastic properties of CNT/cement/stone composites. The accuracy and efficiency of the proposed hierarchical meta-model is demonstrated through detailed parametric analyses. To showcase its potential for optimizing rehabilitation interventions, the paper is closed with its application to a benchmark case study of a stone column injected with CNT/cement grout.
\end{abstract}

\begin{highlights}  
\item Composite CNT/cement grouts are investigated for the first in the literature.
\item A novel nested surrogate modelling approach is proposed for multi-scale composites.
\item Fast stiffness estimates of porous stone injected with CNT/cement grout are obtained.
\item The proposed approach is used to assess the performance of grout injected columns.
\end{highlights}


\maketitle


\section{Introduction}\label{Sect0}

Masonry structures represent an integral part of Europe's architectural heritage and constitute some of the most iconic secular and sacred monumental buildings. {In this light, important research efforts have been devoted to developing structural rehabilitation techniques pursuing the principles of minimal intervention and safeguard of the architectural/artistic values~\cite{buda2022existing,gkournelos2022seismic}. Traditional retrofitting methods include, among others, the addition of shear or infill walls, steel braces, confinements (ring beams or tie bars), transversal anchorage systems, steel reinforced grouts, or grout injections (refer to \cite{yavartanoo2022retrofitting} for a comprehensive state-of-the-art review). In this context, the recent strides in Nanotechnology and Materials Science have unveiled unprecedented opportunities for the development of innovative micro- and nano-engineered materials with vast possibilities across disciplines. Although still largely unexplored, these advancements offer exciting possibilities for creating new minimally invasive rehabilitation techniques endowed with superior mechanical properties.}

There is an increasingly frequent use of composite materials for strengthening and repairing both modern and historic masonry constructions. {Especially in the form of externally bonded composite layers, fiber-reinforced polymers (FRP) such as carbon fibre and glass fibre reinforced polymers (CFRP/GFRP), and textile reinforced mortars (TRM)~\cite{liberotti2022natural,xue2022seismic,cannizzaro2023discrete} have been increasingly used. FRP systems have been rapidly} adopted due to their ease of application, versatility, high tensile strength and stiffness-to-weight ratio, as well as the progressive reduction in their manufacturing and distribution costs. Nonetheless, such systems suffer from durability issues related to debonding, poor performance at high temperatures and sensitivity to ultra violet rays, as well as improper transpiration of the masonry substrate causing compatibility issues~\cite{babatunde2017review}. Another promising field of application of composite materials for rehabilitation of built heritage regards grout injections. Reinforcing grout injections consist of low-pressure (0.5–1 atm) injections of fine cement-based grout through holes drilled into the {masonry~\cite{corradi2008experimental}. This technique represents a widely accepted approach} in rehabilitation~\cite{icomos2003recommendations} for several purposes: to repair and halt the spread of cracks; to increase resistance to moisture penetration; and to strengthen historic structures in seismic areas. These grouts are commonly made of Portland cement (PC) or hydraulic lime as the main binder, aggregates, and additives such as brick dust, epoxy, pozzolana, bentonite, ultra-fine slag, and fly ash or clays~\cite{azadi2017optimization,sha2018experimental}. {However, despite their widely spread use in masonry rehabilitation}, cement-based grouts suffer from some limitations such as low tensile strength and toughness, being susceptible to cracking with the subsequent penetration of water and chemicals accelerating the degradation of the material~\cite{HOGANCAMP2017405}. Recent developments in Nanotechnology have opened vast new possibilities for the development of high-performance composite materials. {Notably}, nanomaterials such as nano calcium carbonate (nano-CaCO$_3$)~\cite{shaikh2014mechanical}, nano silica (nano-SiO$_2$)~\cite{quercia2012water}, graphene and derivatives such as graphene nano-platelets (GnPs) or graphene oxide (GO)~\cite{yee2023review}, and carbon nanotubes (CNTs)~\cite{ramezani2022carbon} have proved exceptional potential to develop high-performance multi-functional cement-based composites (refer to~\cite{csahmaran2022recent} for a comprehensive literature review). Research on cement mortars incorporating CNTs has been particularly copious in the last years, reporting increases of around 10\%, 25\%, and 30\% the compressive strength, bending and Young's moduli of pristine mortar, respectively, by adding low CNT dosages ($\leq1$ wt\%)~\cite{ELGAMAL2017531,LIEW2016301}. However, specific research on the development of nano-modified grouts is not so abundant. One noteworthy contribution from the limited body of existing literature is the one by Liu and co-authors~\cite{liu2021experimental}, who investigated the addition of CNTs for the development of cement composite grouts in inclined fracture grouting with flowing water. Their results demonstrated increases in the sealing efficiency of the composite grout from $\approx$60 to 70\% for CNT contents increasing from 0.5 to 2\%. Another pioneering work was conducted by Lee \textit{et al}.~\cite{lee2019void}, who reported the development of self-heating electrically conductive CNT-cement grouts for the detection of voids in prestressed concrete through thermal imaging and electrical resistance analyses.

The modelling of CNT-based composite materials, while critical for optimizing the composite admixture, poses a formidable challenge given their multi-scale nature~\cite{han2022research}. In the first place, inclusions are defined at the nano-scale, where the filler-matrix load transfer properties are characterized by inter-atomic potentials. At this scale, the most common approaches are~\cite{Rafiee2014}: (i) Molecular Dynamics (MD) simulations; (ii) Continuum modelling; and (iii) atomistic-based continuum approaches. Although MD can analyse high-fidelity representations of the filler-matrix atomic interaction, their application is often limited by computational constraints when analysing systems involving large numbers of atoms. Continuum mechanics-based models can alleviate such limitations by simulating inclusions through continuum structural elements (e.g. beams or shells), although their accuracy may be questionable when adopting concepts from conventional continuum mechanics at the nanoscale~\cite{Chang2006}. Atomistic-based continuum techniques combine the advantages of both methods by properly correlating the interatomic potentials with the stiffness of certain continuum elements such as truss or shell elements~\cite{garcia2019multiscale}. Once the atomic properties of the composites are determined, a micromechanics approach with an appropriate upscaling rule is necessary to estimate the microscopic/macroscopic properties. To this aim, particularly popular approaches are mean-field homogenization (MFH) methods based on the Eshelby's fundamental solution~\cite{Mura1987}. The extension of these methods after the work by Mori and Tanaka~\cite{mori1973average} has proved a computationally efficient solution for simulating large-scale and complex microstructures in an analytical or semi-analytical framework. These techniques can incorporate multiple micromechanical features, {including geometrical characteristics of the inclusions and their orientation distribution, as well as filler waviness, agglomeration, and interphase effects~\cite{giordano2005order,GARCIAMACIAS201849,hassanzadeh2018micromechanical}. For instance, Garc\'{i}a-Mac\'{i}as and Castro-Triguero~\cite{garcia2018coupled} demonstrated, using an extended Mori-Tanaka model, that the interplay between filler waviness and agglomeration leads to coupled weakening effects on the overall stiffness of CNT-composites. Rao \textit{et al.}~\cite{rao2017micromechanics} developed a double-inclusion MFH model to predict the thermo-viscoelastic properties of CNT-reinforced polymers, considering imperfect interfaces. Their findings highlight the significant role of interface effects in influencing the overall properties of CNT-based composites, emphasizing that assuming perfect interfaces leads to overestimates of the stiffness of the composites. These approaches can also be successfully extended to predict other physical properties of composites, including the effective piezoresistive~\cite{buroni2021closed} or piezoelectric~\cite{mishra2022comparative} properties, among others. A noticeable contribution was made by Hassanzadeh-Aghdam and co-authors~\cite{hassanzadeh2018comprehensive} who reported the development of an effective medium (EM) micromechanics-based method to estimate the thermal conductivity of cementitious nanocomposites doped with CNTs, considering filler waviness and interfacial region contributions}. However, the mechanical interaction between fillers is defined through simplifying assumptions, with no consideration of the actual micro-mechanical stress-strain state. To {overcome this limitation, computational homogenization methods offer a more sophisticated yet computationally intensive approach}. These methods consist of simulating a set of virtual tests on a representative volume element (RVE) of the microstructure discretized using numerical simulation approaches such as the finite element (FEM) or the boundary element method (BEM)~\cite{trofimov2018overall,naili2020short}. In this context, it is noteworthy to mention the contribution made by Charitos and co-authors~\cite{charitos2020prediction} who evaluated the effective elastic properties of CNT/polymer nanocomposites by direct 3D FEM. This approach relies on the explicit discretization of the composite phases, requiring advanced mesh strategies and demanding FEM calculations. {As a result, its applicability is limited} to fillers with short to moderate aspect ratios and low volume fractions. {In this context, an embedded element techniques (EETs) have been proposed as a computationally efficient alternative to simulate more complex fiber-reinforced microstructures}. These approaches simulate fillers as 1-D line elements embedded in the matrix through certain constraint conditions. In this line, Lu \textit{et al.}~\cite{lu20143d} developed an EET with a two-step searching \& coupling technique to constraint the movement of fillers inside the matrix. That approach allowed discretizing the matrix phase with a regular non-conforming mesh, enabling the simulation of complex 3D fiber arrangements with large aspect ratios ($\leq 300$) and volume fractions (10-40 wt\%) with limited computational costs. Other advanced homogenization techniques comprise numerical simulation approaches accounting for inclusion-matrix interaction and interfacial effects, Fast Fourier Transform techniques, damage and fracture mechanics, {and multi-scale homogenization aided by Artificial Intelligence}. Interested readers may consult references~\cite{elmasry2022comparative,moud2022recent} for a recent state-of-the-art review.

In light of the literature review conducted above, it is clear that there is great potential for the development of nano-modified cement grouts. In this context, the present work presents a numerical investigation of the possibilities of CNTs to develop high-performance cement-based grouts for the rehabilitation of weak porous stone. This study {builds upon} recent research activities conducted by the authors, {involving} the assessment of the structural condition of heritage constructions erected in Andalusia (Spain) between the 15 and 18$^{\textrm{th}}$ centuries. Many of these monumental buildings were made by the so-called San Crist\'{o}bal stone, including iconic constructions such as the Monastery of San Jer\'{o}nimo de Buenavista~\cite{pozo2003intervencion}, the cathedral of Seville~\cite{jimenez2009proyecto}, or the Santiago church~\cite{padura2011bearing}. The San Crist\'{o}bal stone is characterized by {a} low elastic modulus (4-10 GPa), compressive strength (1.5–2.6 MPa), and high porosity (22.4–27\%)~\cite{baeza2022determining}. As a {consequence}, many of these structures have developed severe cracking patterns, necessitating {multiple} rehabilitation interventions in recent decades. In this light, the idea of producing high-performance CNT-reinforced grouts suggests significant potential for the development of efficient rehabilitation techniques with minimal impact on architectural values. To this aim, a hierarchical multi-scale FEM homogenization approach is developed to estimate the effective elastic properties of porous San Crist\'{o}bal stone doped with CNT/cement grouts. Additionally, to enable the mechanical simulation of large-scale structural elements with minimal computational costs, the numerical homogenization model is bypassed by two nested surrogate models, which represents the main novelty of this work. The effectiveness of the developed approach and the effect of the main properties of the constituent phases are assessed through a {series} of detailed parametric analyses. Finally, the potential of the proposed approach is illustrated through a demonstration example evaluating the modal properties of a full-scale column rehabilitated with grout injections. 


\section{General methodology: hierarchical meta-modelling}\label{Sect01}

\begin{figure}[pos=H]
	\centering
		\includegraphics[scale=0.9]{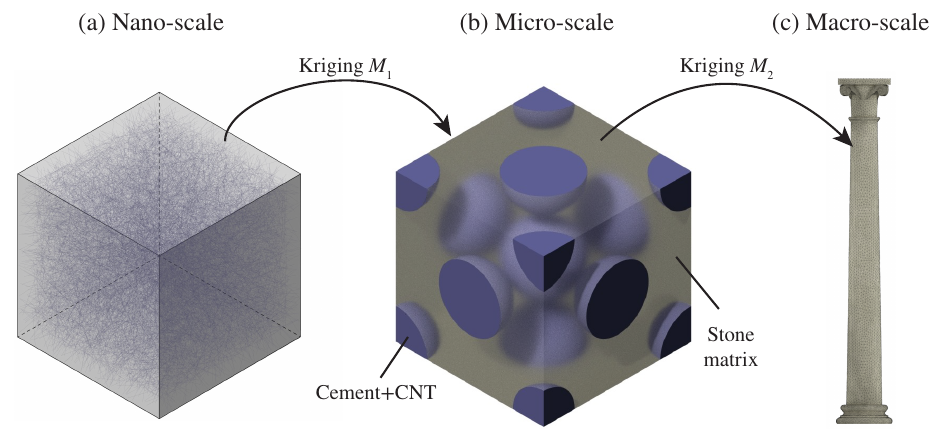}
	\caption{Flowchart of the proposed hierarchical meta-modelling: (a) CNT/cement microstructure, (b) CNT/cement/stone meso-scale, and (c) macro-scale.}
	\label{Flowchart_scales}
\end{figure}

The proposed multi-scale modelling approach is sketched in Fig.~\ref{Flowchart_scales}. It spans across three different scales, including the (a) micro-scale analysing the interaction between CNTs and the cement matrix of the grout, (b) the meso-scale studying the interaction between the matrix stone and the CNT/cement grout occupying the volume of pores and, finally, the (c) macro-scale where the behaviour of complete structural elements is investigated. The effective elastic properties from the first two-scales are estimated through numerical 3D RVEs using FEM. Given the considerable computational burden of such models, which represents a serious difficulty to estimate the response of macroscopic elements, a new hierarchical meta-model is proposed in this work. This involves a set of two nested Universal Kriging models $\mathcal{M}_1$ and $\mathcal{M}_2$. Firstly, meta-model $\mathcal{M}_1$ relates the constitutive properties of the CNTs and the cement matrix with the effective elastic properties of the composite grout. At the meso-scale, the RVE is constituted by spherical inclusions made of CNT/cement composite and a stone matrix. Herein, the effective properties of the grouted pores are estimated using $\mathcal{M}_1$. On this basis, a second meta-model $\mathcal{M}_2$ is constructed to relate the microstructural features of the CNTs, cement, and the stone matrix with the overall elastic properties of the grouted stone. The estimates of this second meta-model are finally used to investigate the macroscopic response of large scale structural elements, and so investigate the effectiveness of the proposed composite grout for rehabilitation purposes.  

\section{Theoretical background}\label{Sect2}

\subsection{Numerical Mechanical homogenization}\label{Sect21}

A classical approach to estimate the effective properties of heterogeneous materials is to adopt a sub-scale modelling approach based on the concept of a RVE. This concept lays on the assumption that any material point $\textbf{X}$ in the continuum scale is associated to a local RVE of volume $\Omega$ and boundary $\Gamma$ (see Fig.~\ref{Scheme_homogenization}). The concept of RVE is only valid in two situations: (i) unit cell in a periodic microstructure: and (ii) volume containing a sufficient number of inclusions to achieve statistical homogeneity and ergodicity in the overall behaviour~\cite{ostoja2002microstructural}. On this basis, in order for the sub-scale-modelling to be energetically consistent, the deformation energy at the macroscopic level must equate to the volume-averaged micro-scale stress work, which is encapsulated by the Hill-Mandel condition~\cite{hill1963elastic}:
 
\begin{figure}[pos=H]
	\centering
		\includegraphics[scale=1.0]{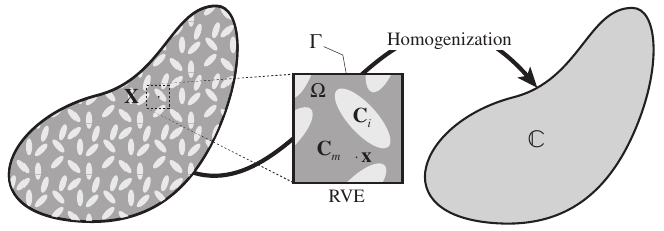}
	\caption{Continuum macro-structure associated to a two-phase RVE consisting of a matrix phase with stiffness tensor $\textbf{C}_m$ and doping inclusions with stiffness tensor $\textbf{C}_i$.}
	\label{Scheme_homogenization}
\end{figure}

\begin{equation}\label{homogenization1}
\overline{\sigma}_{ij}\overline{\varepsilon}_{ij}=\frac{1}{\Omega}\int_\Omega \sigma_{ij} \, \varepsilon_{ij} \textrm{d}\Omega,
\end{equation} 		

\noindent where overbar represents the volume averages of the heterogeneous Cauchy stress tensor $\bm{\sigma}$ and the (virtual) strain tensor $\bm{\varepsilon}$ given by:

\begin{equation}\label{homogenization2}
\overline{\sigma}_{ij} = \frac{1}{\Omega}\int_\Omega \sigma_{ij}\textrm{d}\Omega, \quad 
\overline{\varepsilon}_{ij} = \frac{1}{\Omega}\int_\Omega \varepsilon_{ij}\textrm{d}\Omega.
\end{equation} 		

In the absence of body forces, the Hill-Mandel principle can be rewritten by applying the Gauss theorem as follows~\cite{nguyen2012imposing}:

\begin{equation}\label{homogenization3}
\int_\Gamma\left(t_i-\overline{\sigma}_{ij}n_j\right)\left( u_i- \overline{\varepsilon}_{ik}x_k\right)\textrm{d}\Gamma=0,
\end{equation} 		
	
\noindent where the terms $\textbf{t}$ and $\textbf{u}$ denote the external traction vector and the displacement field on the RVE boundary with normal vector $\textbf{n}$, respectively. On this basis, to fulfil the Hill-Mandel principle, the boundary conditions (BCs) of the sub-scale problem must satisfy Eq.~(\ref{homogenization3}) in order for the homogenization to be energetically consistent. The corresponding Boundary Value Problem (BVP) is typically solved by applying the macro-strain tensor to the RVE. Then, the macro-stress and macro-strain tensors are directly estimated from Eq.~(\ref{homogenization2}) or, alternatively, through surface integration:

\begin{equation}\label{homogenization5}
\begin{split}
\overline{\sigma}_{ij} &= \frac{1}{\Omega}\int_\Gamma t_{i}x_j\textrm{d}\Gamma,\\
\overline{\varepsilon}_{ij} &= \frac{1}{2\Omega}\int_\Gamma \left(u_in_j+u_jn_i\right)\textrm{d}\Gamma.
\end{split}
\end{equation} 		

Once the proper BCs are applied to the RVE, the effective stiffness tensor $\mathbb{C}$ can be estimated from the linear relation between the macrostress tensor $\bm{\overline{\sigma}}$ and the macro-strain tensor $\bm{\overline{\varepsilon}}$ as:

\begin{equation}\label{homogenization4}
\overline{\sigma}_{ij} = \mathbb{C}_{ijkl}\overline{\varepsilon}_{kl}.
\end{equation}

\subsubsection{Nanoscale - CNT/cement composite}\label{Sect211}

\begin{figure}[pos=H]
	\centering
		\includegraphics[scale=0.85]{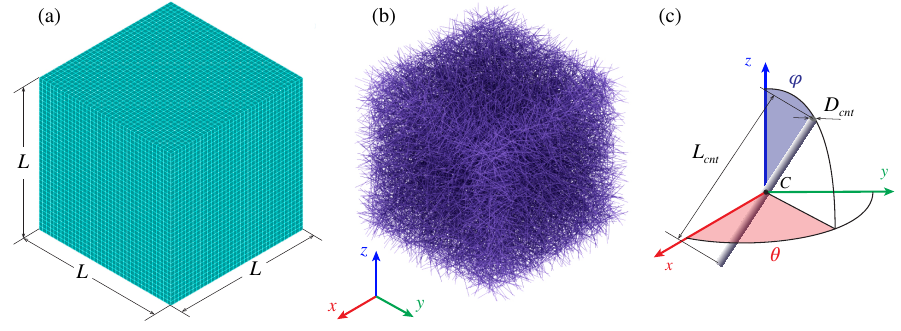}
	\caption{Nano-scale: RVE of cement mortar doped with CNTs. (a) Mortar matrix; (b) CNTs; and (c) Euler angles used to define the orientation of CNTs.}
	\label{RVEs_CNT}
\end{figure}

The first step of the proposed homogenization approach consists of the estimation of the effective properties of the cement mortar doped with CNTs at the nano-scale. Let us consider a cubic RVE of edge length $L$ constituted by a cement matrix doped with randomly oriented CNTs (Fig.~\ref{RVEs_CNT} (a,b)). The RVE is designed to comply with four key assumptions: (i) it contains a sufficiently representative number of CNTs to statistically represent the composite material as a whole; (ii) the length $L_{cnt}$ and diameter $D_{cnt}$ of the CNTs are constant; (iii) the fillers are randomly oriented; and (iv) the mortar matrix can be modelled as a continuum medium. The microstructures are generated through a set of custom-made scripts developed in the MATLAB environment. Each CNT is described by its centre point $C$ as a random variable within the volume of the RVE, and its orientation is defined by means of two Euler angles $\theta$ and $\varphi$ randomly distributed in the intervals $\theta \in \left[0,2\pi\right]$ and $\varphi \in \left[0,\pi/2\right]$, respectively (see Fig.~\ref{RVEs_CNT} (c)). Fillers are added sequentially to the RVE until achieving the desired volume fraction, considering non-intersection conditions between fillers. 

\begin{figure}[pos=H]
	\centering
		\includegraphics[scale=1.0]{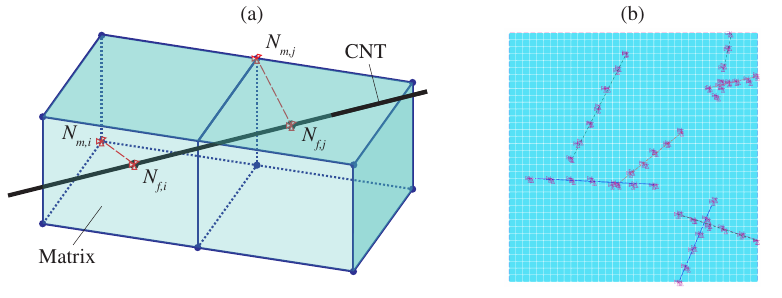}
	\caption{Schematic representation of the ASC algorithm (a), and application to the FEM of a RVE of cement mortar doped with CNTs (b).}
	\label{ASC_sketch}
\end{figure}

Both the matrix phase and the CNTs have been modelled assuming isotropic constitutive properties, as a reasonable simplification commonly adopted in the literature (refer e.g. to~\cite{georgantzinos2022finite,zhang2023influence}). Given the computational difficulties involved in the modelling of large volume fractions of randomly oriented CNTs ({aspect ratio$\approx$1000}) using solid elements, the modified RSA algorithm proposed in~\cite{iorga2008numerical,pan2008analysis} has been adopted in this work. This approach allows modelling fiber-like inclusions by means of computationally light beam elements, and the matrix phase by a non-conforming (planar or three-dimensional) regular mesh (Fig.~\ref{RVEs_CNT} (a)). On this basis, the deformations of the fibers and the matrix are constrained considering perfect interfacial bonding. This alleviates the serious computational demands of complex conforming meshing, resulting in important reductions in the computational time involved in the mesh generation. The algorithm comprises two steps for each fibers' nodes: finding the closest node in the matrix phase using a searching algorithm, then coupling the translational degrees of freedom  (DOFs) of the identified nodes (Fig.~\ref{ASC_sketch}). To compensate the augmented stiffness resulting from the mass matrix covering the volume space that the CNTs should occupy, the Young's modulus of the fibers, $E_f$, is modified as $\hat{E}_f = E_f-E_m$~\cite{lu20143d}, with $E_m$ being the elastic modulus of the matrix phase. 

The resulting RVEs have been meshed and simulated using the FEM code ANSYS19 R3. Specifically, CNTs have been discretized using 2-nodes (6 DOFs per node) 3D Timoshenko beam elements (BEAM188) with circular cross-section, while the mortar matrix has been discretized by a regular array of 8-nodes (3 DOFs per node) solid elements (SOLID185). Given that geometrical periodicity has not been imposed because of the difficulties in achieving high volume fractions when non-intersection conditions among the fibers are considered, the effective elastic properties of the RVE have been estimated from the BVP in Eq.~(\ref{homogenization4}) after adopting linear displacement BCs (Dirichlet). Such BCs are defined to be compliant with the Hill-Mandel principle in Eq.~(\ref{homogenization3}) by imposing a set of displacements $u_i$ at every point $\textbf{x}$ on the boundary $\Gamma$ of the RVE such that:

\begin{equation}\label{KUBC1}
u_i = \overline{\varepsilon}_{ij}x_j, \quad \forall \textbf{x} \in \Gamma.
\end{equation} 	

The stiffness tensor is determined after applying six independent loading conditions (3 pure dilations and 3 pure distortions)~\cite{yan2003homogenization,garcia2022surrogate} to the RVE. For each loading case, one of the six components of the macroscopic strain tensor $\overline{\varepsilon}_{ij}$ is imposed, while the others are set to zero. Afterwards, the stress field in the heterogeneous material is estimated from the static equilibrium analysis, and the average stress field components $\overline{\sigma}_{ij}$ are calculated using Eq.~(\ref{homogenization5}). Next, the effective stiffness tensor $\mathbb{C}$ can be directly derived from a generalised Hooke's law in Eq.~(\ref{homogenization4}). Note that, despite perfectly random orientation distributions of CNTs are well known to lead to perfectly isotropic conditions, the randomness in the construction of the RVE will always induce certain residual anisotropy. This can be minimized through integration over all possible orientations in the Euler space, often referred to as the orientational average. To this aim, the tensor $\mathbb{C}$ can be arbitrarily rotated by the transformation matrix $\bm{a}$ consisting of $\theta$, $\varphi$, and $\psi$ Euler angles as $\mathbb{C}^*_{ijkl}(\theta,\varphi,\psi)=a_{ip}a_{jq}a_{kr}a_{ls}\mathbb{C}_{pqrs}$. On this basis, the orientational average of $\mathbb{C}^{*}$ can be computed as~\cite{schjodt2001mori}:

\begin{equation}
\label{SCH}
\langle \mathbb{C}^{*} \rangle=\displaystyle\frac{1}{4\pi^2}\int_0^{2\pi} \int_0^{2\pi} \int_0^{\pi/2}\mathbb{C}^{*}(\theta,\varphi,\psi)\sin\varphi\,\textrm{d}\theta\,\textrm{d}\varphi\,\textrm{d}\psi.
\end{equation}

\noindent and the effective elastic modulus $\overline{E}$ and the Poisson’s ratio $\overline{\nu}$ can be directly extracted from the compliance tensor $\mathbb{S} = \langle \mathbb{C}^{*} \rangle^{-1}$ (in Mandel-Voigt contracted notation) as:

\begin{equation}
\overline{E} = \dfrac{1}{\mathbb{S}_{11}}, \quad \overline{\nu} = -\overline{E} \, \mathbb{S}_{12}.
\end{equation}

\subsubsection{Microscale - CNT/cement/stone composite}\label{Sect212}

\begin{figure}[pos=H]
  \centering
    \includegraphics[width=\textwidth]{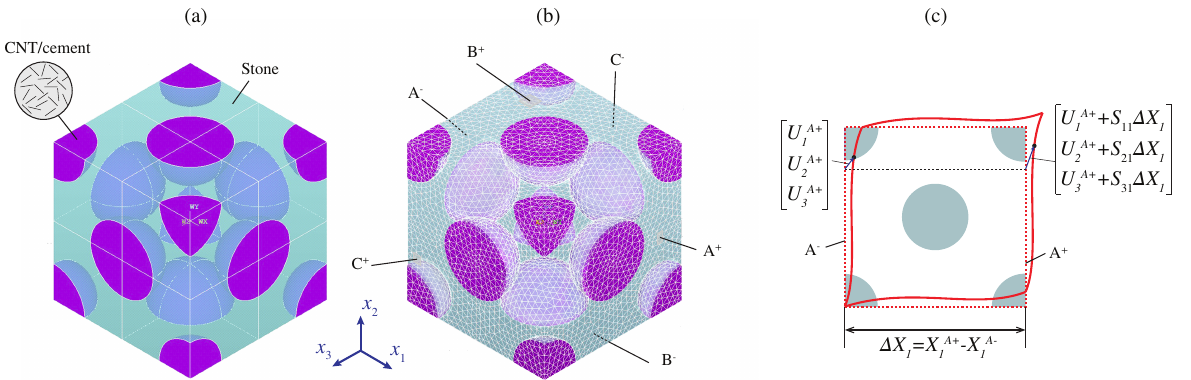}
	\caption{RVE of CNT/cement/stone composite (a), FEM mesh (b) (porosity = 20\%), and schematic representation of periodic BCs (c).}
	\label{Periodic_BCs}
\end{figure}

The second step in the proposed homogenization approach is aimed at estimating the effective constitutive properties of porous stone injected with CNT/cement mortar. It is assumed, for the sake of simplicity, that the pores within the stone are uniformly distributed and possess approximately spherical geometries with a constant diameter. Under these assumptions, as {often adopted in the literature as a first approximation for porous materials (refer e.g. to~\cite{GRASSETBOURDEL20113136,zhang2022relation})}, the RVE of the microscale can be defined through a face-centred cubic cell as shown in Fig.~\ref{Periodic_BCs} (a).

To compute the effective properties of this second RVE, periodic BCs have been applied. To do so, consider the notation of the cell surfaces $A^-/A^+$, $B^-/B^+$, and $C^-/C^+$ indicated in Fig.~\ref{Periodic_BCs} (b). On this basis, the displacements on a pair of opposite boundary cells (with their normal along the $x_j$ axis) read:

\begin{equation}\label{per2}
u_i^{K+}=\hat{\varepsilon}_{ij}x_j^{K+}+v_i^{K+}, \quad u_i^{K-}=\hat{\varepsilon}_{ij}x_j^{K-}+v_i^{K-},
\end{equation}

\noindent with indexes $K+$ and $K-$ indicating the displacements along the positive and negative $x_j$ direction, respectively. The local fluctuations $v_i^{K-}$ and $v_i^{K+}$ should be identical on every two opposite faces to satisfy the periodicity of the RVE. Thus, the local displacements can be dropped from Eq.~(\ref{per2}) by the difference between the two expressions, which leads to:

\begin{equation}\label{per3}
u_i^{K+} - u_i^{K-}=\hat{\varepsilon}_{ij}\left(x_j^{K+}-x_j^{K-}\right).
\end{equation}

The BCs in Eq.~(\ref{per3}) have been implemented in ANSYS through coupling conditions between the displacements of the nodes on opposite boundaries. For each pair of nodes with identical in-plane coordinates (up to a certain tolerance) on two opposite cell faces, the constraint condition among their displacement components is set according to Eq.~(\ref{per3}). By way of example, Figs.~(\ref{Periodic_BCs}) (b) and (c) show the constraint equations to be defined for a pair of nodes located on the cell faces $A^-$ and $A^+$. {Interested readers can} delve into further in-depth details on the implementation of periodic BCs in reference~\cite{Berger2005}. On this basis, likewise the previous model, the components of the effective elastic tensor can be extracted from from Eq.~(\ref{homogenization4}) after adopting six independent virtual tests (3 pure dilations and 3 pure distortions).

\subsection{Surrogate modelling: Kriging models}\label{Sect23}

\begin{figure}[pos=H]
	\centering
		\includegraphics[width=0.93\textwidth]{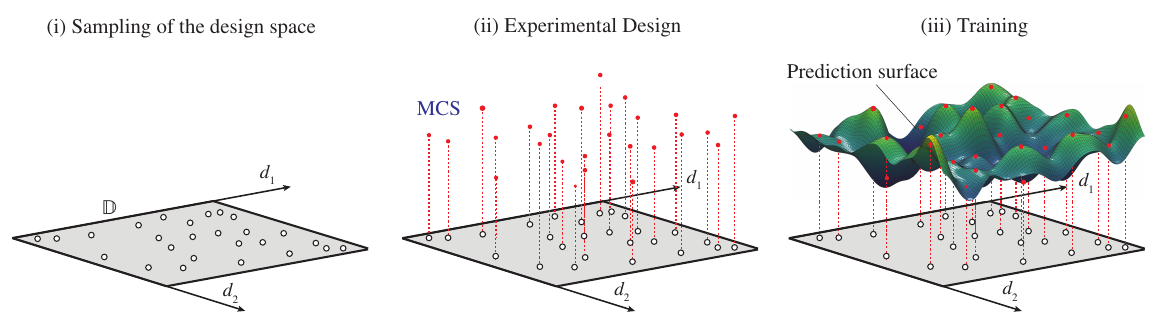}
	\caption{Schematic representation of the work-flow of the construction of a two-variables non-intrusive surrogate model.}
	\label{Sketch_Kriging}
\end{figure}

The estimation of the effective properties using the numerical RVEs outlined in Sections~\ref{Sect211} and \ref{Sect212} is computationally intensive, which limits their implementation for simulating full-scale macrostructural elements. To address this issue, this work proposes the implementation of computationally light surrogate models to bypass the previous numerical homogenization models. Specifically, given the unavailability of the source codes of the utilized commercial FEM software, non-intrusive surrogate models trained on discrete realizations of the forward FEM models are adopted for simplicity. The construction of a non-intrusive surrogate model generally involves three consecutive stages as sketched in Fig.~\ref{Sketch_Kriging} (showing as an example a two-variables model for simplicity in the representation), including: (i) Sampling of the design space; (ii) Generation of the training population through Monte Carlo simulations (MCS) (also referred to as the experimental design (ED)); and (iii) Construction of the prediction surface. 

Let us consider $m$ design variables $d_i \in \mathbb{R}, i=1,\ldots ,m$ allowed to vary only within a certain physically meaningful range $\left[a_i,b_i\right]$. Accordingly, the vector of design variables $\textbf{d}=\left[d_1,\ldots ,d_m \right]^\textrm{T}$ spans the $m$-dimensional design space $\mathbb{D}=\left\{{\textbf{d} \in \mathbb{R}^m:a_i \leq d_i \leq b_i }\right\}$. To train the surrogate model, it is necessary to assemble a training population of $N_s$ individuals mapping the output $y$ (quantity of interest) and the design space $\mathbb{D}$, also referred to as the ED. This is achieved by uniformly sampling the input design space $\mathbb{D}$ to conform a matrix of design sites $\textbf{D}=[\textbf{d}^1,\ldots,\textbf{d}^{N_s} ] \in \mathbb{R}^{(m \times N_s)}$. The corresponding outputs $y^i$ are then obtained by direct Monte Carlo simulations (MCS) of the main FEM (in this case, the homogenization of the RVEs) and collected in an observation vector $\textbf{Y}=\left[y^1,\ldots,y^{N_s} \right]^\textrm{T}$.

In this study, among the wide range of non-intrusive surrogate models available in the literature, the Kriging model is chosen due to its remarkable adaptability to a large variety of problems~\cite{Kleijnen2009}. The Kriging interpolator represents the function of interest $y(\textbf{d})$ by combining a regression term $y_r \left(\textbf{d}\right)$ and a zero-mean stochastic process $\mathcal{Z}\left(\textbf{d}\right)$ as~\cite{Kleijnen2017}:

\begin{equation}\label{Kriging1}
y\left(\textbf{d}\right)=y_r \left(\textbf{d}\right)+ \mathcal{Z}\left(\textbf{d}\right). 		
\end{equation}

The regression function $y_r (\textbf{d})$ depends upon $p$ linear regression parameters $\bm{\kappa}=\left[\kappa_1,\ldots,\kappa_p\right]^\textrm{T}$ and certain user-defined regression functions $f(\textbf{x})=\left[f_1 (\textbf{d}),\ldots,f_p (\textbf{d})\right]^\textrm{T}$ with $f_i:\mathbb{R}^m \rightarrow \mathbb{R}$ as:

\begin{equation}\label{Kriging2}
y_r (\textbf{d})=f(\textbf{d})^\textrm{T} \bm{\kappa}.
\end{equation}

The stochastic process $\mathcal{Z}\left(\textbf{d}\right)$ can be characterized by a covariance function $\textrm{Cov}\left[\mathcal{Z}(\textbf{d}_i)\mathcal{Z}(\textbf{d}_j)\right]$ between any two data points $\textbf{d}_i$ and $\textbf{d}_j$:

\begin{equation}\label{Kriging3}
\textrm{Cov}\left[\mathcal{Z}(\textbf{d}_i)\mathcal{Z}(\textbf{d}_j)\right]=\sigma^2 r\left(\textbf{d}_i,\textbf{d}_j,\bm{\theta} \right),  		
\end{equation}

\noindent where $\sigma^2$ denotes the variance of $\mathcal{Z}\left(\textbf{d}\right)$, and $r\left(\textbf{d}_i,\textbf{d}_j,\bm{\theta} \right)$ is a user-defined spatial correlation function dependent on $\bm{\theta}$ hyper-parameters. Once defined, the Kriging model predicts the response $y \left(\textbf{d}\right)$ at an arbitrary design site $\textbf{d}$ as:

\begin{equation}\label{Kriging4}
\widehat{y}(\textbf{d})=f(\textbf{x})^\textrm{T} \bm{\kappa}+r(\textbf{d})^\textrm{T} \textbf{R}^{-1} \left[ \textbf{Y}-f(\textbf{d})^\textrm{T} \bm{\kappa} \right],
\end{equation}

\noindent where $\textbf{R}$ is a $N_s \times N_s$ correlation matrix with components $R_{ij}= r\left(\textbf{d}_i,\textbf{d}_j,\bm{\theta} \right)$, with $r(\textbf{d})$ being a vector of correlations between the all the samples in the DS and $\textbf{d}$, that is:

\begin{equation}\label{Kriging5}
r(\textbf{d})^\textrm{T}=\left[ r\left(\bm{\theta},\textbf{d}_1,\textbf{d}\right),\ldots,r\left(\bm{\theta},\textbf{d}_{N_s},\textbf{d}\right)\right]^\textrm{T}.  	
\end{equation}

From Eq.~(\ref{Kriging4}), it is clear that the Kriging interpolator is fully defined by the regression parameters $\bm{\kappa}$ and the correlation parameters $\bm{\theta}$. We adopt second-order polynomials for the trend term, and Gaussian correlation functions for the stochastic term as~\cite{Sacks1989}:

\begin{equation}\label{Kriging6}
r\left(\textbf{d}_i,\textbf{d}_j,\bm{\theta} \right) = \prod_{k=1}^{m} \exp\left[-\theta_k \left(d_i^{(k)}-d_j^{(k)} \right)^2\right].	
\end{equation}

For given values of the hyper-parameters $\bm{\theta}$, the trend parameters $\bm{\kappa}\left( \bm{\theta} \right)$ and the variance $\sigma^2 \left( \bm{\theta} \right)$ can be optimally computed in closed-form by means of the empirical best linear unbiased estimator (BLUE) (consult \cite{Kleijnen2017} for more extensive information). On the other hand, the estimation of the hyper-parameters $\bm{\theta}$ involves a non-linear optimization problem. In this work, the maximum-likelihood-estimator is solved through the iterative pattern search optimization algorithm implemented in the DACE toolbox for Matlab developed by Lophaven and co-authors~\cite{Lophaven2002}. Finally, note that hyper-parameters $\theta_k$ in Eq.~(\ref{Kriging6}) dictate the shape of the correlation function, allowing for potential anisotropy along the dimensions of $\textbf{d}$. However, for the sake of simplicity, correlations are assumed isotropic in this work, that is $\theta_k=\theta \quad \forall \, 1 \leq k \leq m$.

\section{Numerical results and discussion}\label{Sect3}

The numerical results and discussion of the presented hierarchical meta-modelling approach for fast estimation of the elastic properties of CNT/cement/stone composites are presented in this section. The material parameters employed in the subsequent analyses are furnished in Table~\ref{material_properties}. It is important to remark that the values in this table cover the range of the most commonly used bases in grout injections, spanning from lime to cement mortars (elastic modulus from 2 to 10 GPa, and Poisson's ratios from 0.05 to 0.3). 

\begin{table}[pos=H]							
\newcommand\Tstrut{\rule{0pt}{0,3cm}}         
\newcommand\Bstrut{\rule[-0.15cm]{0pt}{0pt}}   
\footnotesize							
\caption{Constitutive properties of the considered phases to investigate CNT/cement/stone composite materials.}				
\vspace{0.1cm}							
\centering							
\begin{tabular}{cccccc}							
\hline																																					
Quantity	&	\multicolumn{5}{c}{Material}			\Tstrut\Bstrut\\
\cline{2-6}	
	                    &	MWCNTs	          &	&	Cement mortar	    &	&	Stone	\Tstrut\Bstrut\\			
\hline	
Young's modulus	      &	0.23-1.22 TPa$^a$	&	&	2-10 GPa$^c$	    &	&	4-10 GPa$^e$	\Tstrut\\
Poisson's ratio [-]	  &	0.1-0.28$^b$	    &	&	0.05-0.3$^d$	    &	&	0.3	\\
Diameter [nm]	        &	1.8	              &	&	-	                &	&	-	\\
Length [µm]	          &	0.41	            &	&	-	                &	&	-	\\
Volume fraction [\%]	&	0.01-1.2	        &	&	-	                &	&	-	\\
Porosity [\%]	        &	-	                &	&	-	                &	&	5-30$^f$	\Bstrut\\
\hline	
\multicolumn{6}{l}{\footnotesize Sources: $^a$\cite{el2018computational}; $^b$\cite{zhao2011study}; $^c$\cite{drougkas2019confinement}; $^d$\cite{drougkas2019confinement}; $^e$\cite{baeza2022determining} ; $^f$\cite{baeza2022determining}} \Tstrut\\																				
\end{tabular}		
\label{material_properties}							
\end{table}	

The results hereafter have been organised into validation analyses in Section~\ref{Sect3.1}, construction of the surrogate models in Section~\ref{Sect3.2} and, finally, a toy example of a stone column to illustrate the potential of the developed approach for the design of composite injections in Section~\ref{Sect3.3}.

\subsection{Validation analyses}\label{Sect3.1}

The definition and accuracy of the RVE built for CNT/cement composites is appraised first. A critical aspect of computational homogenization techniques regards the size of the RVEs as extensively discussed in the literature (see e.g.~\cite{chen2019average,bai2023determination}). {It is well known that when fibers are randomly oriented, their strengthening effects distribute equally in all directions of space, resulting in isotropic stiffness tensors (refer e.g. to reference~\cite{garcia2018coupled}}). Therefore, the determination of the critical size relies on statistical theory, wherein the departure from isotropy is evaluated as a function of the RVE size. To analyse the deviation of the effective stiffness tensor from ideal isotropy, we adopt the anisotropy index~\cite{zerhouni2019numerically}:

\begin{equation}\label{homogenization}
\delta_{iso} = \frac{\left|\mathbb{C}-\langle \mathbb{C}^{*} \rangle \right|_{F}}{\left| \mathbb{C} \right|_{F}},
\end{equation} 

\noindent where $\left| \mathbb{A} \right|_{F} =  \sqrt{\textrm{tr}{(\mathbb{A}\cdot \mathbb{A}^\textrm{T})}}$ denotes the Frobenius norm of a fourth-order tensor $\mathbb{A}$. On this basis, a perfectly isotropy tensor leads to an anisotropy index $\delta_{iso}=0$.

\begin{figure}[pos=H]
	\centering
		\includegraphics[width=1\textwidth]{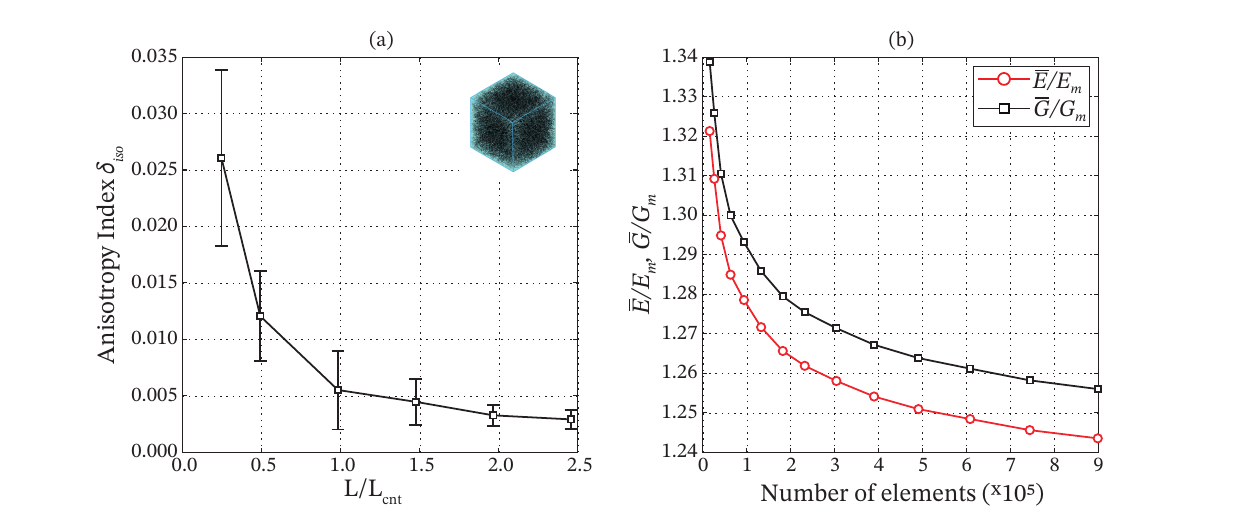}
	\caption{Departure from isotropy versus the RVE size (a) mesh convergence (b) for the RVE of CNT/cement composites ({filler volume fraction of 0.5\%}). The error bars in (a) represent the standard deviation of the anisotropy indexes obtained after five realizations.}
	\label{Convergence_validation_model_1}
\end{figure}

Figure~\ref{Convergence_validation_model_1} (a) reports the convergence analysis conducted to determine the critical size of the RVE. To do so, the size of the RVE for CNT/cement mortars is defined in a dimensionless way as $L/L_{cnt}$ ranging between 0.25 and 2.46. Specifically, 6 different RVEs are generated with 5 realizations each to analyse the dispersion of the estimated effective properties. {In these analyses, the filler volume fraction is fixed to 0.5\%} and the filler and matrix phases are meshed considering element sizes of $L_{cnt}/6$ and $L_{cnt}/18$, respectively. It can be observed in this figure that convergence is approximately achieved at $L/L_{cnt}=1.96$  $(L=0.80 \; \upmu \textrm{m})$. Beyond this limit, the residual anisotropy remains approximately constant and, therefore, the RVE size $L/L_{cnt}=1.96$ is fixed in the subsequent analyses. 

It is equally important to study the sensitivity of the homogenization results to discretization density of the RVE to ensure the estimates are mesh-independent. To do so, different mesh densities have been investigated for the same RVE ($L=0.80 \; \upmu \textrm{m}$, CNT volume fraction 0.5\%), ranging between 15,417 and 899,148 elements. These discretizations correspond to different mesh densities used in the study. The mesh sizes for the filler and matrix phases of the RVE range from $L_{cnt}/3$ to $L/9$ and from $L_{cnt}/16$ to $L/48$, respectively. Figure~\ref{Convergence_validation_model_1} (b) furnishes the convergence analysis conducted in terms of the normalized effective elastic moduli $\overline{E}/E_m$ and $\overline{G}/G_m$ versus the number of finite elements, $E_m$ and $G_m$ being respectively the elastic and shear moduli of the matrix phase. From the results, it is concluded that convergence is approximately achieved for a RVE discretization of 744,291 elements, which corresponds to a mesh density of $L_{cnt}/15$ and $L/45$ for the filler and matrix phases, respectively.

\begin{figure}[pos=H]
	\centering		
	\includegraphics[width=1\textwidth]{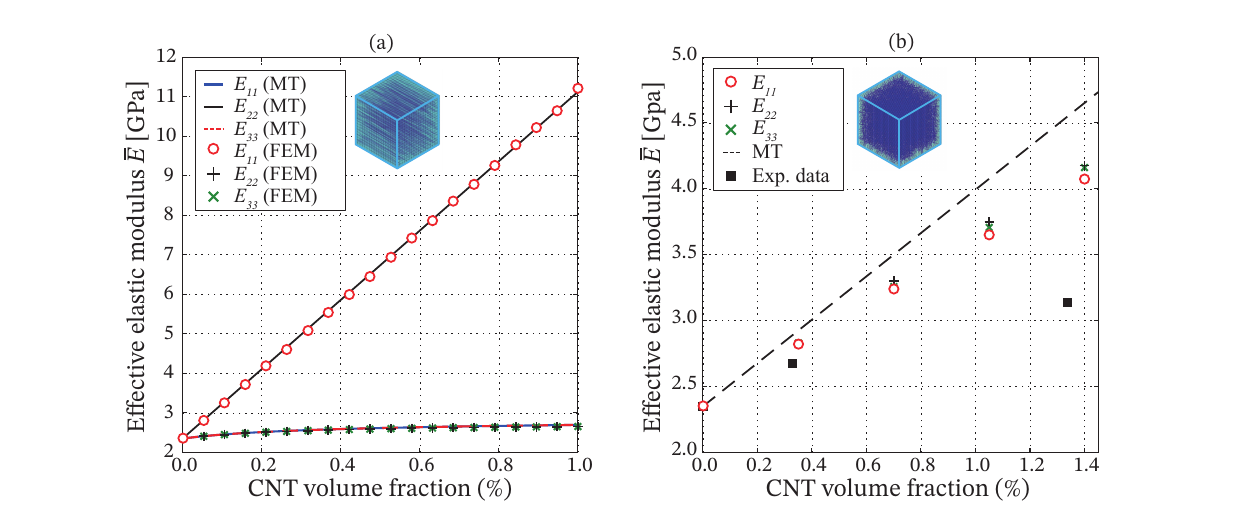}
	\caption{Validation analysis of the numerical estimates of the elastic properties of PC/CNT composites considering fully aligned (a) and randomly oriented filler configurations (b) (experimental data from~\cite{talo2017updated}), {and (c) comparison between the numerical/analytical predictions of the elastic modulus for against experimental data for cement/CNT (experimental data from~\cite{kavvadias2023mechanical})}.}
	\label{Validation_aligned_random}
\end{figure}

The accuracy of the resulting RVE is investigated in Fig.~\ref{Validation_aligned_random} by comparing the predictions of the computational homogenization with the estimates obtained using the classical Eshelby-Mori-Tanaka (MT) MFH approach, as well as the experimental data reported by {Tal{\`o} and co-authors~\cite{talo2017updated} and Kavvadias \textit{et al.}~\cite{kavvadias2023mechanical} for polymer/CNT and cement/CNT composites, respectively. In the analyses in Figs.~\ref{Validation_aligned_random} (a) and (b),} the investigated material corresponds to polycarbonate (PC) thermoplastic polymer reinforced with single-walled CNTs (SWCNT) ($L_{cnt}=1.3 \; \upmu \textrm{m}$, $D_{cnt}=1.8 \; \textrm{nm}$). Both phases are assumed isotropic, with elastic moduli of 2.35 GPa and 0.97 TPa for the PC and SWCNT, respectively, and Poisson's ratios of 0.38 and 0.1. Two different filler configurations are investigated, namely fully aligned and randomly oriented fillers in Figs.~\ref{Validation_aligned_random} (a) and (b), respectively. In the case of fully aligned fillers in Fig.~\ref{Validation_aligned_random} (a), 20 different RVEs ($L=0.20 \; \upmu \textrm{m}$) with CNT volume fractions ranging from 0 to 1\% and perfectly aligned in the $x_1$ direction (i.e.~$\theta=0$ and $\varphi=\pi/2$) have been constructed and compared with the predictions of the MT model. In this case, the composites exhibit transversely isotropic properties, {with a considerable stiffer behaviour in the direction of the fibers ($E_{22}=E_{33}<E_{11}$). It is noted in this case that} close agreements are found between the numerical and the analytic predictions. In the case of randomly oriented fillers in Fig.~\ref{Validation_aligned_random} (b), five different RVEs have been calculated ($L=0.80 \; \upmu \textrm{m}$), and the numerical predictions are benchmarked against both the analytical MT solution and the experimental data. It is noted that both the estimates obtained by the MT model and the numerical homogenization are close to the experimental data for low CNT volume fractions. Nevertheless, considerable differences arise with the increase in the volume fraction of CNTs. Note, however, that the numerical estimates are more proximate to the experimental data, while the MT model leads to considerable overestimates of the effective properties. {The differences between the experimental data and the theoretical estimates can be attributed to microstructural aspects disregarded in the homogenization, such as the effects of filler agglomeration and waviness, as well as imperfect filler-matrix load transfer interactions. These phenomena have been reported in the literature to act as microstructural defects, consequently weakening the effective properties of the composite (refer e.g. to references~\cite{talo2017updated,hassanzadeh2019creep,zhu2018effect}). Notably, it has been reported in the literature that as the filler content increases and CNTs become closer to each other, the likelihood of forming filler rises~\cite{yeh2006mechanical}, which is in line with the results presented in Fig.~\ref{Validation_aligned_random} (b). Moreover, agglomeration effects can lead to increasing non-linear correlations between the effective elastic modulus and the CNT content~\cite{garcia2018coupled}, which further supports the need for including these microstructural features in future developments. In the analyses shown in Fig.~\ref{Validation_aligned_random} (c), the elastic moduli of cement and CNTs are selected as 11.30 GPa and 1.0 TPa, respectively, and the filler aspect ratio is set to 157.89 ($L_{cnt}=1.5\; \upmu \textrm{m}$ and $D_{cnt}=9.5$ nm) following reference~\cite{kavvadias2023mechanical}. The filler content is presented in this figure as mass fraction (wt), with the bulk mass densities of cement and CNTs chosen as 1440 and 0.18 g/cm$^3$, respectively. Note that in this case the numerical results underestimate the Young's modulus in in all cases (maximum relative error of 13.53\% for wt=0.1\%), except for wt=0.3\%. In this case, very close agreements are found, with slight overestimates by the numerical model. Importantly, these errors are within the same order of magnitude as those reported in similar numerical homogenization models previously documented in the literature (e.g.~\cite{kavvadias2023mechanical,al2012aspect}). Alongside the aforementioned factors, these differences may also be attributed to CNT-induced alterations in the hydration process of cement and the effects of ultrasonic sonication on the CNT aqueous solution~\cite{ramezani2021elastic}}. Nevertheless, since the aim of this investigation is to develop a methodology suitable for conducting a preliminary assessment of the potential of CNT/cement composites for rehabilitation grouts, the accuracy of the presented formulation is deemed sufficient for this purpose. The incorporation of further microstructural details is left for a future work.

\begin{figure}[pos=H]
	\centering
	\includegraphics[width=1\textwidth]{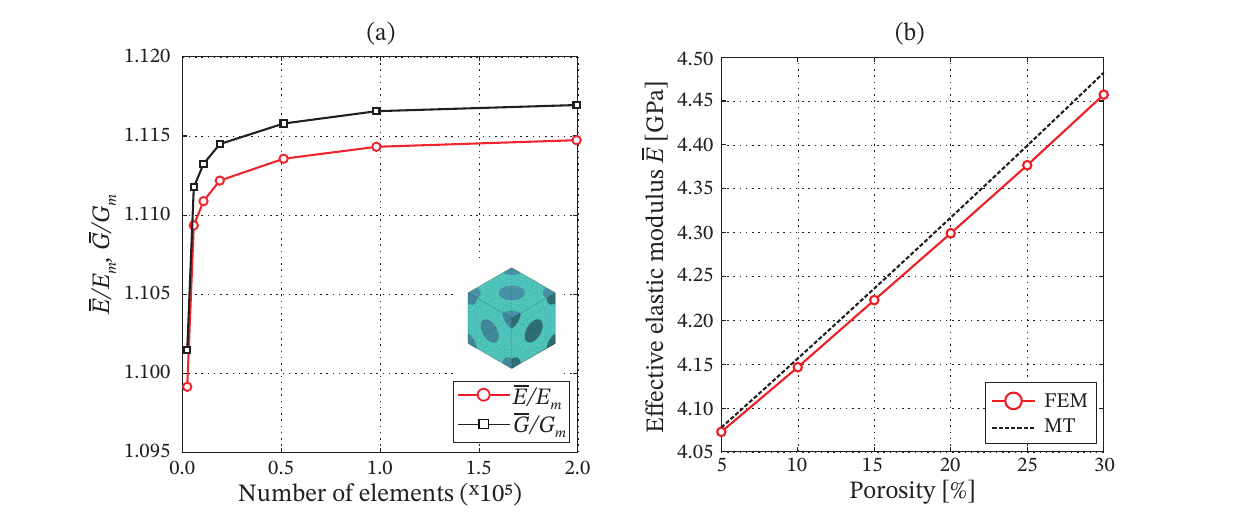}
	\caption{Convergence analysis for the RVE of CNT/cement/stone composites (a) and validation against the theoretical estimates by MT (b) ({1\% CNT volume fraction}, 500 GPa and 0.1 the elastic modulus and Poisson's ratio of CNTs, 5 GPa and 0.3 the Young's modulus and Poisson's ratio of cement).}
	\label{Validation_metamodel_2}
\end{figure}

Finally, the validation analyses of the RVE developed for CNT/cement/stone composites are reported in Fig.~\ref{Validation_metamodel_2} (a). In these analyses, the elastic moduli of the CNTs, cement, and stone are kept fixed at 0.5 TPa, 5 GPa, and 4 GPa, respectively. Additionally, the Poisson's ratios of the CNTs, cement, and stone are also maintained constant with values of 0.1, 0.3, and 0.3, respectively, {and the CNT volume fraction is set at 1.0\%}. Firstly, the convergence of the mesh density of the RVE is investigated in Fig.~\ref{Validation_metamodel_2} (a). In this case, one single RVE accounting for a porosity of 30\% (maximum considered value) and a size of 0.5 microns is defined and discretized using seven different mesh densities with a total number of elements from  2,328 to 199,352. To achieve mesh convergence, it is determined that a mesh density of 784,384 elements per unit volume is required for this RVE. Afterwards, the accuracy of the selected mesh density is benchmarked against the predictions obtained using the classical MT method in Fig.~\ref{Validation_metamodel_2} (b). The results in this figure reveal that both approaches yield very similar results, although the predictions by the MT method slightly overestimates those obtained by FEM as previously reported in the literature (see e.g.~\cite{mortazavi2013modeling}).

\subsection{Surrogate modelling}\label{Sect3.2}

\begin{figure}[pos=H]
	\centering		           
	\includegraphics[width=1\textwidth]{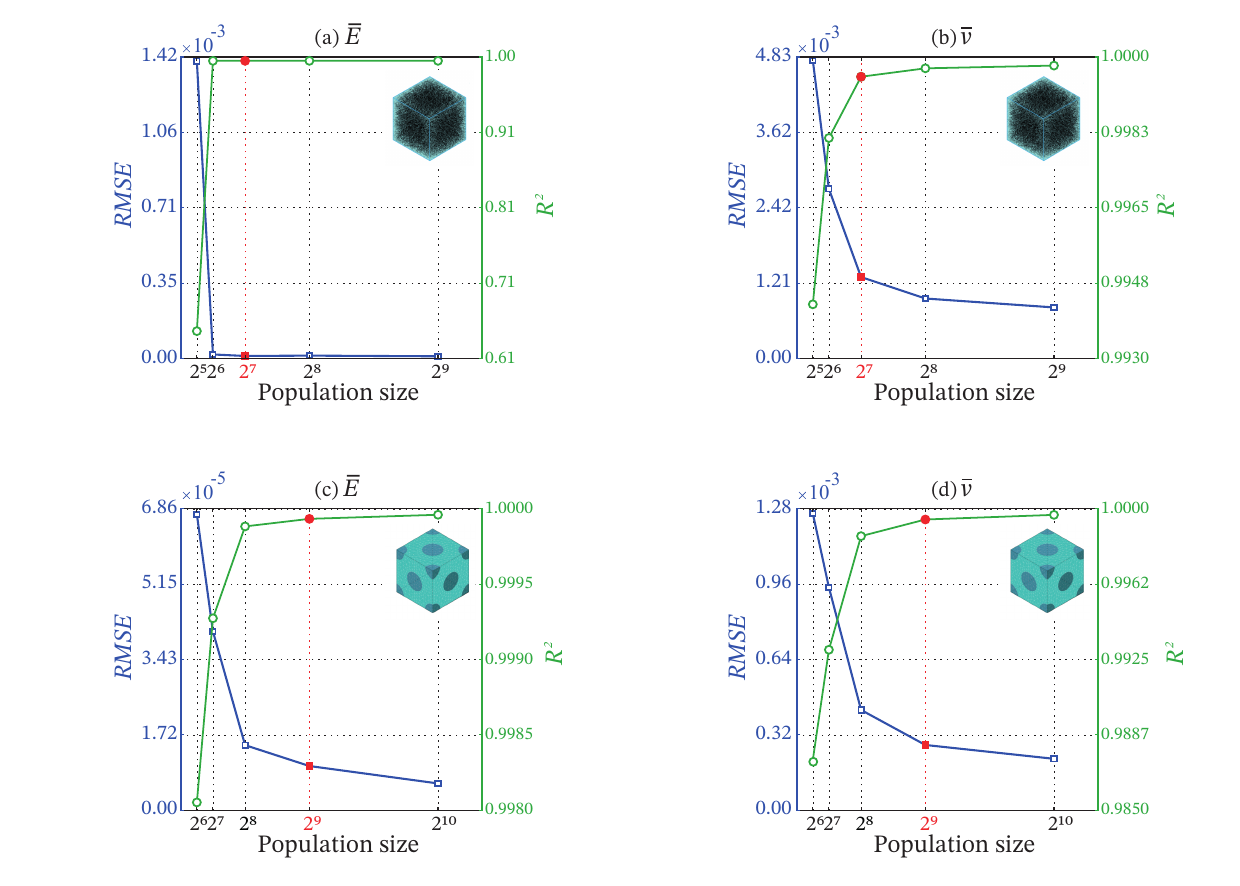}
	\caption{Convergence analysis of metamodels $\mathcal{M}_1$ (a,b) and $\mathcal{M}_2$ (c,d) in terms of effective elastic modulus (a,c) and Poisson's ratio (b,d) random for increasing training populations.}
	\label{convergence}
\end{figure}

The first surrogate model, $\mathcal{M}_1$, considers five different input variables, namely the CNTs volume fraction, the elastic modulus and Poisson's modulus of CNTs, and the elastic modulus and Poisson's ratio of the cement matrix. On the other hand, the second surrogate model, $\mathcal{M}_2$, adds two additional input variables, comprising the porosity and Young's modulus of the stone. The variation ranges of these parameters defining the design space $\mathbb{D}$ have been set according to the ranges previously reported in Table~\ref{material_properties}. The accuracy of any surrogate model is largely determined by the quality of the ED. The quantity and distribution of training samples is highly case-dependent, often necessitating specific analyses to adapt the ED to the variability of the quantity of interest. Typically, it is necessary to achieve a uniform sampling of the design space to cover the entire domain of interest. To this aim, Latin Hypercube Sampling has been adopted in this work to draw uniformly distributed samples over $\mathbb{D}$. To ascertain the appropiate size of the ED, convergence analyses have been performed considering varying training populations of increasing size. For $\mathcal{M}_1$, five different training populations have been considered, ranging from 32 to 512 individuals, with a validation set (VS) of 1024 samples. For $\mathcal{M}_2$, six training populations have been defined, ranging from 32 to 1024 individuals, with a VS of 2048 samples. For every population, the effective Young's modulus and Poisson's ratio of each individual are obtained by forward evaluation of the numerical RVEs previously outlined in Section~\ref{Sect21}. Note that, likewise any other non-intrusive meta-model model bypassing a computationally intensive model, the generation of the training population represents the most computationally intensive step. Then, the convergence of the accuracy of the predictions by the surrogate-model is evaluated in Fig.~\ref{convergence} in terms of the Root Mean Square Error (RMSE) and the coefficient of determination R$^2$, measured over the VS, between the predictions of the surrogate models and the forward FEM models. As expected, Fig.~\ref{convergence} shows that the convergence rate of $\mathcal{M}_2$ is slower than that of $\mathcal{M}_1$ due to the larger number of input variables. Based on these results, the convergence of meta-models $\mathcal{M}_1$ and $\mathcal{M}_2$ is approximately achieved for population sizes of 2$^7$ and 2$^9$ individuals, respectively. In these cases, RMSE values close to zero and R$^2$ coefficients close to one ensure the high performance of the surrogate models. 

\begin{figure}[pos=H]
	\centering
		\includegraphics[width=1\textwidth]{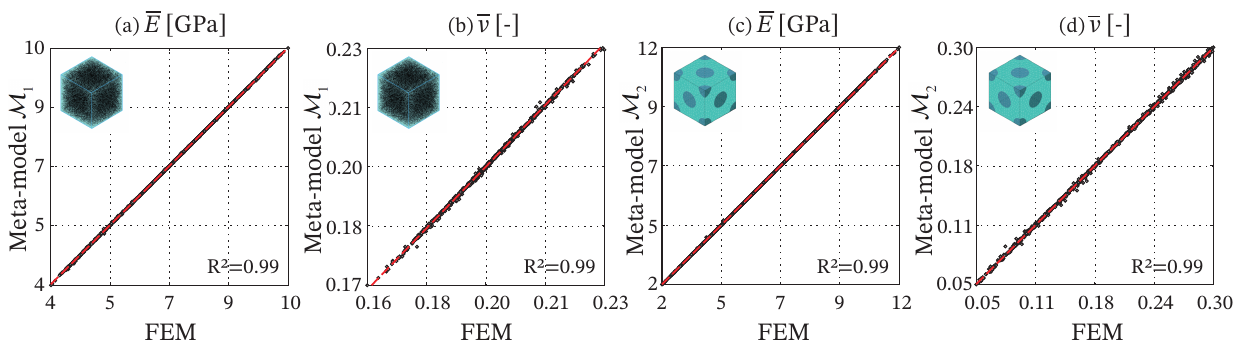}
	\caption{Scatter plots of the predictions by the Kriging meta-models $\mathcal{M}_1$ (a,b) and $\mathcal{M}_2$ (c,d) versus the forward evaluations of the FEM RVEs for the VSs of 1024 and 2048 samples.}
	\label{Scatter_plots}
\end{figure}

The surrogate models $\mathcal{M}_1$ and $\mathcal{M}_2$, trained with 2$^7$ and 2$^9$ individuals respectively, are selected for the subsequent analyses. Figure~\ref{Scatter_plots} depicts the comparison between the predictions of the forward RVEs and the resulting surrogate models. These plots, commonly used in surrogate modelling applications, provide a visual representation of the goodness-of-fit between the predictions of the meta-models and the forward FEM models. The minimal dispersion of the data points along the diagonal lines (R$^2$ values very close to 1) in Fig.~\ref{Scatter_plots} provides strong evidence that the surrogate models are trained with high precision. Both meta-models also exhibit low computational burden, with mean evaluation times of 0.18 seconds and 0.25 seconds for meta-models $\mathcal{M}_1$ and $\mathcal{M}_2$, respectively. These reduced computational evaluation times make the meta-models an efficient solution for conducting intense parametric analyses, iterative optimization, as well as the analysis of full-scale structural elements.

\subsection{Parametric analyes}\label{Sect3.3}

\begin{figure}[pos=H]
	\centering
		\includegraphics[width=1\textwidth]{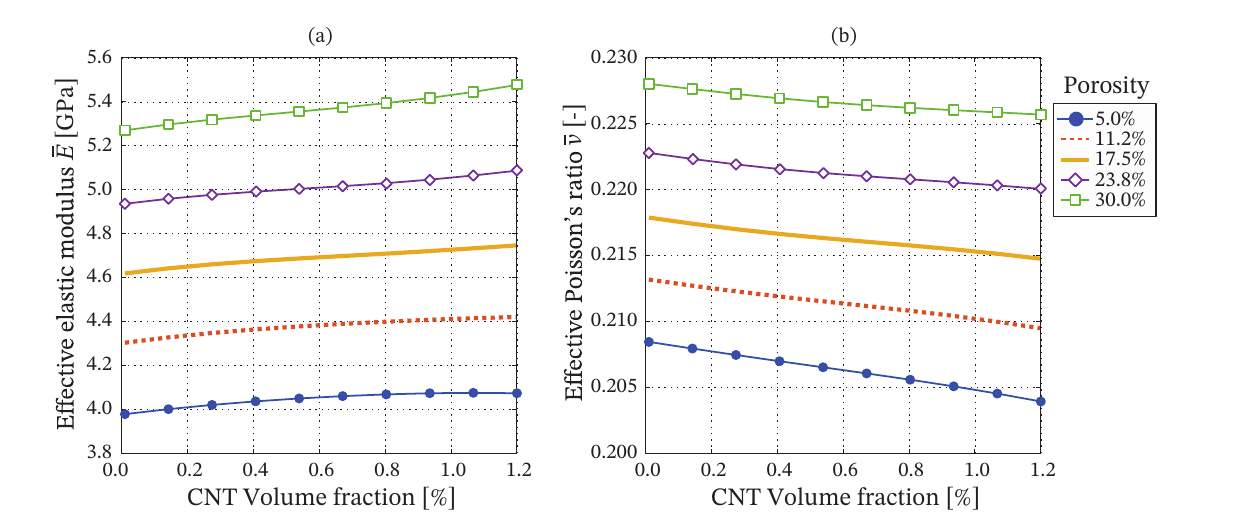}
	\caption{Variation of the effective elastic modulus (a) and Poisson's ratio (b) of CNT/cement/stone composites versus the CNT volume fraction for varying porosity percentages in stone.}	
	\label{Code Parametric case 1}
\end{figure}

\begin{figure}[pos=H]
	\centering
		\includegraphics[width=1\textwidth]{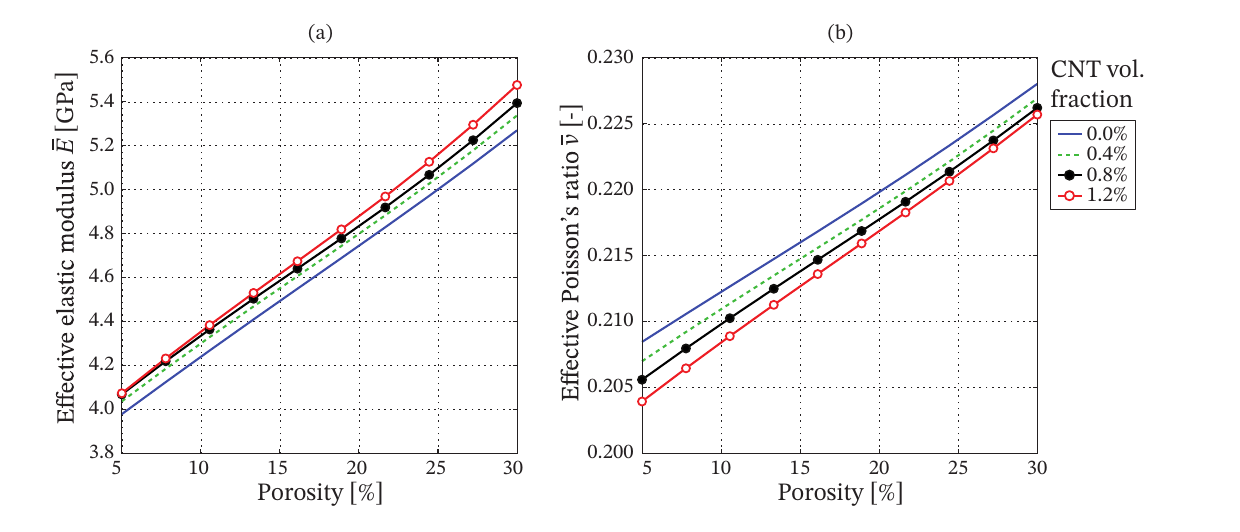}
	\caption{Variation of the effective elastic modulus (a) and Poisson's ratio (b) of CNT/cement/stone composites versus the porosity content for varying CNT volume fractions.}	
	\label{Code Parametric case 2}
\end{figure}

In order to showcase the potential of the proposed nested meta-models, a comprehensive set of parametric analyses evaluating the impact of the input variables on the effective elastic properties of CNT/cement/stone composites is presented herein. The effects of the CNT volume fraction and the porosity in stone are first investigated in Fig.~\ref{Code Parametric case 1}. In these analyses, the elastic moduli of the CNTs, cement, and stone are kept fixed as 1.0 TPa, 10 GPa and 4 GPa, respectively, {and the} Poisson's ratio of the CNTs and cement are also maintained as constant with values of 0.1 and 0.3, respectively. {Furthermore, the CNT volume fraction is systematically varied from 0.0 to 1.2\%}, and 5 levels of porosity are considered in the stone ranging from 5\% to 30\%. On this basis, the variations of the elastic modulus and Poisson's ratio of the CNT/cement/stone composites are presented in Figs.~\ref{Code Parametric case 1} (a) and (b), respectively. It is noted in Fig.~\ref{Code Parametric case 1} (a) that the effective elastic modulus monotonically increases as so does the CNT volume fraction, with more pronounced gradients as the porosity increases. {From a physical point of view, this indicates that a greater amount of CNTs in the composite grout occupying the porosity in the stone contributes to the overall rigidity of the grouted stone}. The opposite behaviour can be observed in Fig.~\ref{Code Parametric case 1} (b) for the effective Poisson's ratio, which is ascribed to the comparatively lower Poisson's ratio of the CNTs compared to the San Crist\'{o}bal stone (see Table~\ref{material_properties}). Interestingly, the Poisson's ratio increases for higher porosity levels, which is conceivably explained by the larger content of cement grout occupying the porosity and with a similar ratio to stone's. These analyses are complemented with a second parametric analysis in Fig.~\ref{Code Parametric case 2}. This figure presents the analysis of the effects of the porosity level in stone on the overall elastic properties of CNT/cement/stone composites. Specifically, the porosity is systematically varied within the range from 5 to 30\%, as typically observed in the San Cristóbal stone. Furthermore, five cases of CNT volume fractions are considered from 0\% to 1.2\%. The results in Fig.~\ref{Code Parametric case 2} (a) reveal an increase in the effective elastic modulus with increasing porosity, this trend being more acute for a higher CNT volume fractions. Similarly, the results in Fig.~\ref{Code Parametric case 2} (b) reveal increases in the effective Poisson's ratio as the porosity increases. These results agree with the previously reported ones in Fig.~\ref{Code Parametric case 1}, finding lower ratios for higher CNT volume contents. 

\begin{figure}[pos=H]
	\centering
		\includegraphics[width=1\textwidth]{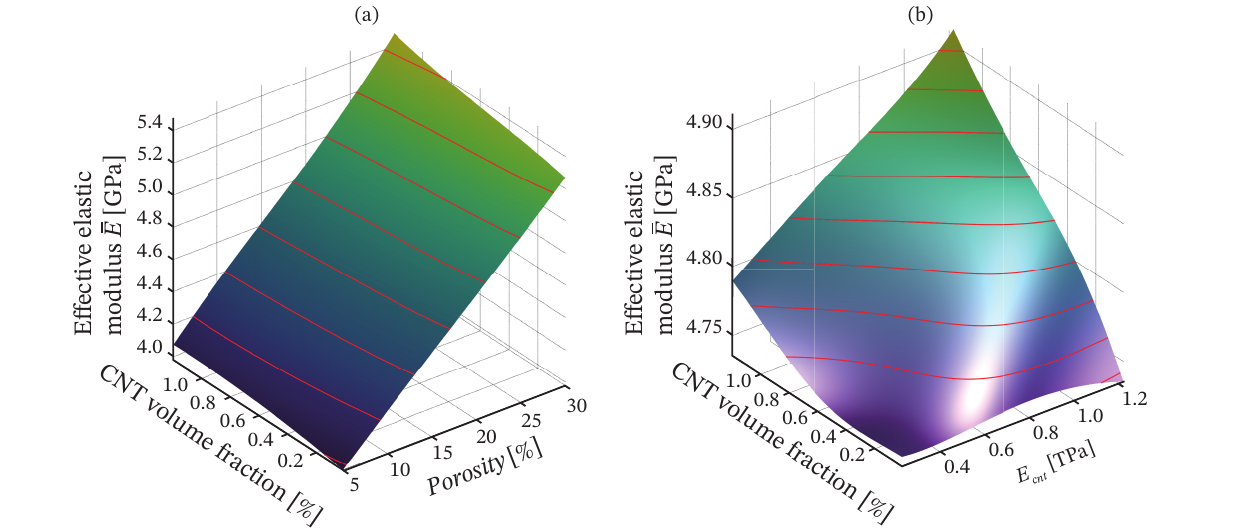}
	\caption{Effective elastic modulus of CNT/cement/stone composites as a function of the CNT volume fraction and porosity (a), and CNT volume fraction and elastic modulus (b) (porosity=20\%).}	\label{Code Parametric case 3}
\end{figure}

Finally, Fig.~\ref{Code Parametric case 3} summarizes the conducted parametric analyses by presenting the surfaces of the effective elastic moduli of CNT/cement/stone composites as functions of the CNT volume fraction and porosity of the stone (a), and versus the CNT volume fraction and Young's modulus (b). These figures demonstrate that the potential of CNT/cement grouts for increasing the rigidity of stone increases as so does the elastic modulus of CNTs and the porosity of the original stone, achieving maximum increases of around 35\%. These results are reasonable since, as the porosity of the stone increases, so does the volume occupied by the composite grout. In addition, the stiffness of the grout increases as so does the elastic modulus and the volume fraction of the CNTs. Nonetheless, it is important to remark the simplifying assumptions adopted in this work. On one hand, we are not considering filler agglomeration effects, which may have a more significant influence as the CNT volume fraction increases. {Extensive literature reports have consistently documented that CNTs, due to their large specific surface area, tend to form agglomerates induced by van der Waals attraction forces. This gives rise to CNT clusters that act as microstructural defects, which adversely impact the overall effective mechanical properties of the composite~\cite{hassanzadeh2019creep,hassanzadeh2021evaluating}}. Another important aspect disregarded in this work {includes} the connectivity of the pores and permeability of the stone. These factors critically determine the penetration potential of the grout and, therefore, the proportion of pores that can potentially be occupied by the composite grout. While the experience of the authors demonstrates that the San Crist\'{o}bal stone typically presents highly interconnected porosity, making the assumptions adopted in this work reasonable, specific studies in this regard are required to support these analyses. Due to space constraints and the considerable fundamental developments required for such studies, the extension of the presented framework to accommodate filler agglomeration effects and partially grouted pores is left for future work. 

\subsection{Application case study: injected stone column}\label{Sect3.4}
 
\begin{figure}[pos=H]
	\centering
		\includegraphics[width=0.95\textwidth]{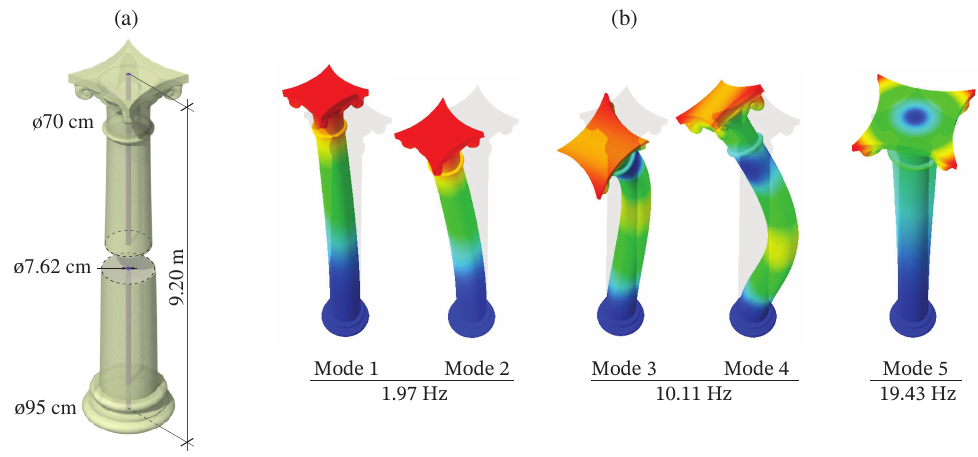}
	\caption{Benchmark case study: injected stone column (a); reference modes of vibration of the non-injected column (b).}
	\label{Toy_example}
\end{figure}

\begin{figure}[pos=H]
	\centering
		\includegraphics[width=1\textwidth]{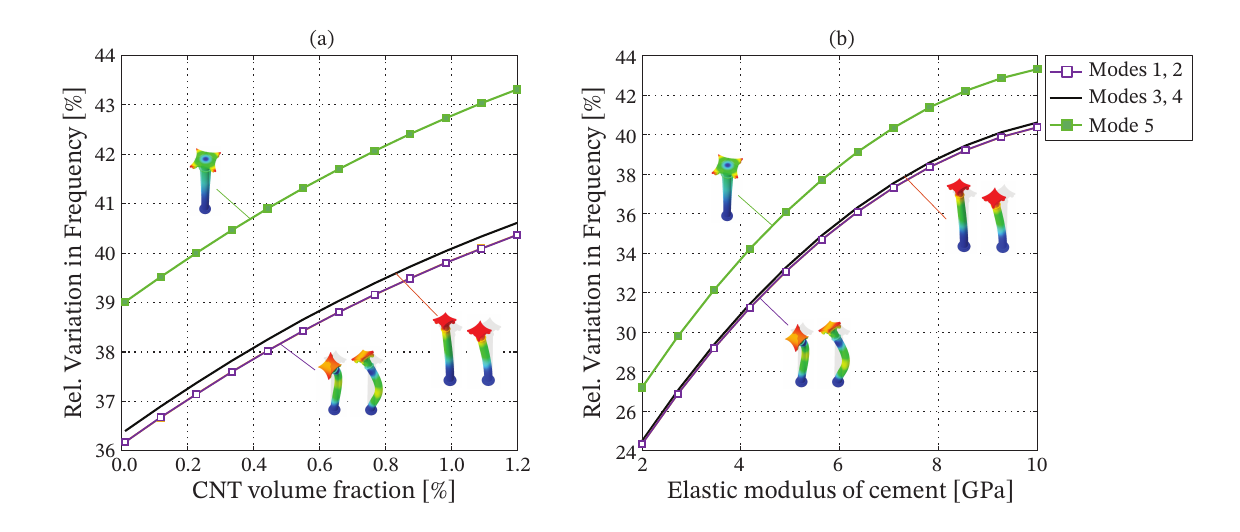}
	\caption{Parametric analysis of the natural frequencies of vibration of the toy column injected with CNT/cement grout considering increasing CNT volume fractions (elastic modulus of cement equal to 10 GPa) (a) and elastic modulus of the cement for a CNT content of 1.2\% (b) (the porosity of the stone is fixed to 20\%)}
	\label{Analysis Parametrico}
\end{figure}

This final case study {is intended to} illustrate the potential of the proposed methodology for the macroscopic analysis of full-scale structures. The example consists of a hypothetical column inspired by one of the columns located in the Alameda de H\'{e}rcules square in Seville. Its dimensions include a total height of 9.20 m, a diameter of 95 cm at the base and 70 cm under the capital. As a potential rehabilitation intervention, this section investigates the effect of injecting the column with CNT/cement grout along its with a perforation of 7.62 cm of diameter as sketched in Fig.~\ref{Toy_example} (a). To this aim, the constitutive properties of the grout and the CNT/cement/composite are estimated using the developed meta-models $\mathcal{M}_1$ and $\mathcal{M}_2$. The properties of the stone used in the analysis include a Young's modulus value of 4 GPa, a Poisson's ratio of 0.3, an an porosity of 20\%. It is assumed that the perforation is filled with CNT/cement grout ($\mathcal{M}_1$), while the remaining section of the column is fully saturated by the injection ($\mathcal{M}_2$). The column has been modelled and discretized in the FEM software ABAQUS using 4-nodes (Q3D4) tetrahedron solid elements (total number of 308,407 elements and 60,387 nodes). Cantilever boundary conditions are applied by restraining only the translational degrees of freedom at the base of the column. On this basis, a set of parametric modal analyses are performed to evaluate the stiffening effects of the composite grout in terms of the natural frequencies of vibration of the column. The variations are quantified as relative changes in the natural frequencies compared to the non-injected configuration of the column (Fig.~\ref{Toy_example} (b)). The analysis includes the first five natural modes: two first order bending modes (symmetric along the main directions of the column), two second order bending modes, and one first order torsional mode. Figure~\ref{Analysis Parametrico} (a) furnishes the relative variations of the resonant frequencies for CNT filler contents ranging from 0 to 1.2\%. It is noted in this figure that the use of standard cement grouts (CNT volume fraction equal zero) leads to an increase in the fundamental frequency of approximately 36\%. Then, when considering small contents of CNTs (1.2\% the volume of cement), this increase can rise up to around 40\%. It is also interesting to investigate the effect of the elastic modulus of cement as reported in Fig.~\ref{Analysis Parametrico} (b). This figure demonstrates that the stiffness of the cement plays a critical role, with the fundamental frequency increasing from 24\% to 40\% as the elastic modulus of the cement varies between 2 to 10 GPa. {It is important to emphasize that the whole parametric analysis, facilitated by the developed surrogate model, were completed in just 27 minutes. In contrast, approaching this problem via a numerical multi-scale approach would present overwhelming, and possibly unfeasible, computational challenges.} Overall, these results highlight the potential of the develop approach for designing rehabilitation interventions using CNT-reinforced cement grouts. They also show great promise as a foundational framework for incorporating complex factors such as filler agglomeration and grout injection permeability.

\section{Concluding remarks}\label{Sect4}

We have presented in this work an innovative hierarchical meta-modelling approach for the rapid prediction of the effective elastic properties of stone injected with CNT/cement grouts. The developed methodology utilizes two nested Kriging meta-models to bypass the numerical homogenization of the two main scales involved in CNT/cement/stone composites with minimal computational burden. Specifically, the nano-scale is defined by a two-phase RVE comprising a cement matrix doped with randomly arranged CNTs. At the micro-scale, a second RVE is defined to represent porous stone injected with CNT/cement grout. The meta-models are trained in a non-intrusive fashion through direct Monte Carlo simulations of the numerical homogenization of the RVEs. Specifically, the first meta-model, $\mathcal{M}_1$, considers as independent variables the CNT volume fraction, as well as the Young's moduli and Poisson's ratios of the matrix and the CNTs (5 variables). The second meta-model, $\mathcal{M}_2$, expands upon $\mathcal{M}_1$ by adding the porosity and the Young's modulus of the stone as additional independent variables. The dimension and mesh density of the RVEs, as well as the sizes of the training populations for the surrogate models, are selected through convergence and statistical analyses. The soundness and effectiveness of the proposed hierarchical meta-modelling approach have been appraised through thorough parametric analyses. The paper has been concluded with a toy example of a stone column, demonstrating the effectiveness of the investigated CNT/cement grouts for structural rehabilitation. The main discoveries of this work can be succinctly summarized as follows:

\begin{itemize}
	\item The conducted convergence analyses have revealed that number of training points required for achieving accurate predictions by the meta-model is directly related to the number of independent input variables. Specifically, the convergence for meta-model $\mathcal{M}_1$ (5 variables) was achieved for $2^7$ individuals, while meta-model $\mathcal{M}_2$ (7 variables) required $2^9$ samples. Notably, the meta-models demonstrated high accuracy and very low computational demands, with evaluation times of 0.18 and 0.25 seconds for meta-models $\mathcal{M}_1$ and $\mathcal{M}_2$, respectively. 
	\item The reported parametric analyses have showcased the effect of key microstructural features upon the elastic moduli of CNT/cement/stone composites, namely the CNT volume fraction, the elastic properties of CNT, cement and stone, as well as the porosity in the stone. The presented numerical results have revealed the potential increase of up to 35\% the elastic modulus of porous stone (30\% of porosity) with a CNT volume fraction of 1.2\% the volume of cement in the grout.
	\item The presented toy example of a 3D stone column has demonstrated the potential of CNT/cement grout for rehabilitation purposes. To this aim, the hierarchical meta-modelling approach presented this study offers a fast and accurate method for optimizing such interventions.
\end{itemize}

{It is important to emphasize that several simplifying assumptions have been considered in the definition of the microstructures for the sake of simplicity in the construction of the nested surrogate models, which constitute the actual innovative contribution of this work. In particular, the inclusion of additional microstructural features, such as filler waviness, agglomeration, and non-uniform filler aspect ratio and orientation distributions should be considered. These additions would enhance the model's accuracy and capture more realistic behaviours. Moreover, accounting for the micro-constituent phases and the porosity of cement, and utilizing more realistic geometrical models for the porosity of stone may significantly improve the accuracy of the proposed model. A potentially valuable extension within the proposed methodology involves the incorporation of non-linear load-bearing links between the CNTs and the matrix phase, with the purpose of predicting the ultimate strength properties of the CNT-reinforced grouts. Finally, analysing the permeability of stone to grout injections would be valuable for a comprehensive optimization of the proposed grout technology. Nonetheless, given the general nature and flexibility of the proposed hierarchical meta-modelling approach, its practical applications and potential for further development render it a promising methodology in the field of composite materials. Overall, the proposed approach provides a rapid and accurate method for predicting the effective elastic properties of stone injected with CNT/cement composite grout, enabling the structural analysis of full-scale elements while incurring minimal computational costs.}


\section*{Acknowledgements}
This work was supported by the Operational Programme 2014-2020 of the European Regional Development Fund and the Consejería de Economía, Conocimiento, Empresas y Universidad of the Andalusian Regional Government (Spain) under project US-1381164. The financial support is gratefully acknowledged. R. R-R was also supported by a FPU contract-fellowship from the Spanish Ministry of Education Ref: FPU18/05211.

\bibliographystyle{unsrt}
\bibliography{bibliography_Clean}
\end{document}